\documentclass[conference]{IEEEtran}
\IEEEoverridecommandlockouts
\usepackage{cite}
\usepackage{amsmath,amssymb,amsfonts}
\usepackage{algorithmic}
\usepackage{graphicx}
\usepackage{textcomp}
\usepackage{xcolor}
\usepackage{tikz}
\usetikzlibrary{shapes.geometric, arrows}
\def\BibTeX{{\rm B\kern-.05em{\sc i\kern-.025em b}\kern-.08em
    T\kern-.1667em\lower.7ex\hbox{E}\kern-.125emX}}

\usepackage{float}
\usepackage{caption}
\begin{document}

\title{Meursault as a Data Point\\
}

\author
{\IEEEauthorblockN{Abhinav Pratap}
\IEEEauthorblockA{\textit{Department of Computer Science} \\
{\textit{and Engineering, ASET}} \\
\textit{Amity University, Noida, India} \\
TheAPratap@gmail.com}
}

\maketitle

\begin{abstract}
In an era dominated by datafication, the reduction of human experiences to quantifiable metrics raises profound philosophical and ethical questions. This paper explores these issues through the lens of Meursault, the protagonist of Albert Camus' The Stranger, whose emotionally detached existence epitomizes the existential concept of absurdity. Using natural language processing (NLP) techniques including emotion detection (BERT), sentiment analysis (VADER), and named entity recognition (spaCy)-this study quantifies key events and behaviors in Meursault's life. Our analysis reveals the inherent limitations of applying algorithmic models to complex human experiences, particularly those rooted in existential alienation and moral ambiguity. By examining how modern AI tools misinterpret Meursault's actions and emotions, this research underscores the broader ethical dilemmas of reducing nuanced human narratives to data points, challenging the foundational assumptions of our data-driven society. The findings presented in this paper serve as a critique of the increasing reliance on data-driven narratives and advocate for incorporating humanistic values in artificial intelligence.
\end{abstract}

\begin{IEEEkeywords}
Datafication, Existentialism, The Stranger, Absurd Hero, Phenomenology, Ethics of AI, Human Data Representation, Postmodernism
\end{IEEEkeywords}

\section{Introduction}
In the digital age, the quantification of human experience has become a dominant paradigm, promising objectivity and predictive power [5]. However, this reductionist approach, known as datafication, risks obscuring the complexity and nuance inherent in human existence [6]. The field of digital humanities grapples with this tension, exploring the affordances and limitations of computational methods in understanding human culture and expression [8]. Albert Camus' The Stranger offers a poignant exploration of this paradox [1]. Meursault, the novel's protagonist, embodies the absurd hero, a figure alienated from societal norms and conventional morality [1][2]. His emotionally detached responses to significant life events, such as his mother's death and his own impending execution, challenge traditional notions of meaning and understanding [1].

This paper seeks to examine whether Meursault's existential experiences can be meaningfully captured through data. Using a dataset generated from The Stranger and analyzed with various NLP techniques, including emotion detection (BERT) [9], sentiment analysis (VADER) [10], and absurdity tagging, we quantify Meursault's behaviors and emotional responses. This empirical approach allows us to explore the ethical and philosophical implications of reducing complex human experiences to structured data [12]. Specifically, we ask: Can data-driven models capture the essence of existential alienation [3]? What are the limitations and ethical concerns of applying modern AI tools to such narratives [13]?

By analyzing Meursault's life as a dataset, we aim to highlight the risks of oversimplification and the loss of meaning in datafication [5], ultimately advocating for a more humanistic approach to data interpretation.

\section{Literature Review}
The intersection of existentialism and data analysis presents a compelling paradox: Can the subjective, nuanced, and often irrational aspects of human existence be effectively captured through the objective, quantifiable lens of data? This literature review explores this question by examining the character of Meursault from Albert Camus' The Stranger [1] and drawing on existentialist philosophy, particularly Camus' concept of absurdity [2]. By synthesizing existing literature and considering the emerging field of digital humanities [8], this review delves into the implications and ethical challenges of reducing the human experience to data [5]. It critically evaluates the limitations and potential risks of datafication, questioning whether the complexity of existential struggles and the human condition can truly be represented in such a reductive framework.
\subsection{Existentialism and the Absurdity of Human Existence}
Central to existentialist philosophy is the belief that human life is inherently absurd—a concept articulated by Albert Camus in The Myth of Sisyphus [2] and The Stranger [1]. According to Camus, the absurd arises from the conflict between humanity's innate desire for meaning and the universe's indifference to this deLsire [2]. Meursault, the quintessential absurd hero, confronts this tension with detachment and indifference, rejecting societal expectations of emotion and morality [1]. His refusal to conform underscores the existential themes of alienation [3], the search for authenticity in a meaningless world [4], and the individual's freedom to create meaning in the face of absurdity [2].

This philosophical stance directly challenges the reductionist mindset prevalent in modern technology. Datafication, the process of transforming human experiences into quantifiable data, operates under the assumption that all facets of human behavior can be measured, analyzed, and understood through data. However, as Camus emphasizes, human existence is marked by subjective and often irrational experiences that defy categorization [1][2]. This inherent resistance to quantification underscores the limitations of data-driven approaches in capturing the complexity of existential themes. Moreover, it raises questions about what is lost when we reduce the richness of lived experiences to mere data points.

\subsection{The Limitations of Data-Driven Reductionism in the Digital Humanities}
The field of digital humanities explores the use of computational methods to analyze and interpret cultural artifacts, including literary texts. While these methods offer new possibilities for understanding human expression, they also raise concerns about the limitations of data-driven reductionism.

Datafication promises to provide objective insights into human behavior, but it often fails to account for the nuances and ambiguities of lived experience. Luciano Floridi argues that while data can simulate patterns of human behavior, it cannot replicate the subjective depth and context that define human existence [5]. This critique is particularly relevant when examining a character like Meursault, whose actions and emotions are deeply rooted in existential absurdity [1].

Our dataset, generated from The Stranger, exemplifies this challenge. For instance, emotion detection models like BERT [9] classify Meursault's reactions as sadness or anger, failing to capture the philosophical indifference     that defines his character. Similarly, sentiment analysis tools such as VADER [10] apply polarity scores to Meursault's narrative, reducing his existential reflections to simplistic positive or negative values. These misclassifications highlight the risk of oversimplification and distortion inherent in datafication. They also underscore the need for more nuanced computational approaches that can better account for the complexities of human experiences, especially those rooted in existentialism.

\subsection{The Ethics of Datafication and the Misinterpretation of Meursault}
The ethical implications of datafication extend beyond literary analysis, raising broader concerns about privacy, bias, and autonomy. Shoshana Zuboff warns of the dangers of ``surveillance capitalism,'' where personal data is commodified and used to manipulate behavior [6]. In this context, the reduction of individuals to data points can lead to dehumanization and loss of agency [12].

Applying these ethical concerns to Meursault's case, we see a parallel in how his identity is reduced to quantifiable metrics. For example, our named entity recognition (NER) model identifies character interactions but cannot capture the existential significance of these relationships. Meursault's detachment from his mother's death is recorded as an absence of emotion, but this reduction ignores the deeper philosophical critique of societal norms embedded in his response [1].

Real-world applications of AI and big data demonstrate similar ethical dilemmas. Predictive policing algorithms, for instance, rely on historical data that often reflect existing biases, leading to discriminatory outcomes [13]. Similarly, hiring algorithms may reinforce gender or racial inequalities if trained on biased datasets [12]. These examples underscore the need for ethical frameworks that prioritize human dignity and complexity over algorithmic efficiency.

This philosophical challenge is not unique to literature. For instance, predictive policing systems often misinterpret calm behavior as suspicious due to inherent biases. Similarly, AI tools in mental health assessments may misclassify emotional detachment as depression. These real-world misinterpretations echo the limitations observed in analyzing Meursault’s narrative, emphasizing the need for AI systems capable of understanding contextual and philosophical nuances.

\subsection{The Need for a Philosophy of Big Data in Literary Studies}
As big data continues to shape our understanding of the world, it is essential to develop a philosophical framework that addresses its epistemological and ethical implications, especially in the context of literary analysis. A robust philosophy of big data must grapple with questions about the nature of knowledge, truth, and human agency in the digital age. Floridi suggests that data-driven technologies must be guided by principles that respect the inherent ambiguity and subjectivity of human experience [5].

In the context of The Stranger, this philosophical framework highlights the limitations of using data to understand Meursault's existential journey. His actions and emotions cannot be reduced to binary classifications or sentiment scores without losing their deeper meaning. By examining Meursault as a data point, we expose the inadequacy of datafication in capturing the richness of human existence and advocate for a more nuanced, ethical approach to data interpretation. This approach should acknowledge the limitations of current AI tools while remaining open to the potential of future advancements in the field.

\section{Conceptual Framework}
This study operates at the intersection of existential philosophy and datafication, using the character of Meursault from Albert Camus' The Stranger [1] as a case study to explore the limitations of reducing complex human experiences to quantifiable data. The framework integrates existentialist concepts with modern AI and NLP techniques, providing a theoretical basis for understanding the ethical and philosophical challenges posed by datafication.

\subsection{Existential Absurdity and Human Experience}
Albert Camus' philosophy of the absurd [2] highlights the inherent conflict between humanity's desire for meaning and the universe's indifference. In The Stranger [1], Meursault epitomizes this absurd heroism through his indifference to societal norms and emotional expectations. This inherent conflict presents a fundamental tension between the subjective nature of human experience, as emphasized by existentialism [3], and the objective, quantifying nature of datafication [5].
\begin{itemize}
    \item Philosophical Challenge:
        \begin{itemize}
            \item Subjectivity vs. Objectivity: Existentialism emphasizes the subjective, ambiguous, and often irrational aspects of human life [3]. These elements resist reduction to objective data points, challenging the fundamental premise of datafication.
            \item Meursault as a Case Study: Meursault's actions and emotional responses—or lack thereof—such as his detachment at his mother's funeral, illustrate the complexities of human experience that defy simple categorization [1]. His existential indifference cannot be neatly classified as sadness, anger, or apathy within a data-driven framework.
        \end{itemize}
\end{itemize}

\subsection{Datafication and Its Reductionist Paradigm}
Datafication involves converting complex human experiences and behaviors into structured, analyzable data. This process assumes that all aspects of life can be measured and understood through quantifiable metrics.
\begin{itemize}
    \item Reductionism: Reducing human experiences to data points risks oversimplifying the richness and ambiguity of lived reality. Luciano Floridi argues that data, while useful for identifying patterns, cannot capture the full depth of subjective experiences [5].
    \item Misrepresentation: Applying datafication to Meursault's narrative demonstrates these limitations. For instance, BERT emotion detection [9] may classify his existential detachment as sadness or anger, failing to account for the philosophical significance of his indifference. VADER sentiment analysis [10] might apply a negative polarity score to his reflections, misinterpreting them within a binary emotional framework.
\end{itemize}

\subsection{Ethical Considerations in AI and Datafication}
\begin{itemize}
    \item Dehumanization: Shoshana Zuboff's concept of surveillance capitalism [6] highlights how personal data can be commodified, reducing individuals to mere data points and stripping away their autonomy and complexity.
    \item Bias and Misinterpretation: AI systems often reflect the biases embedded in their training data, leading to distortions in how human behavior is analyzed and represented [12][13]. This is particularly problematic when dealing with existential themes, which are inherently subjective and resistant to algorithmic interpretation.
    \item Loss of Nuance: The dataset's classification of Meursault's behaviors and interactions reveals how data-driven models fail to capture the philosophical essence of his character. For example, named entity recognition (NER) identifies character interactions but cannot convey the existential significance of these relationships.
    \item Broader Implications: These misrepresentations mirror real-world ethical dilemmas, such as the risk of misclassifying human behavior in areas like criminal justice or mental health, where oversimplification can lead to harmful outcomes.
\end{itemize}

\subsection{Bridging Existentialism and Data Science}
\begin{itemize}
    \item Ambiguity and Meaning: Existentialism asserts that life's meaning is not inherent but must be created by individuals [3]. This subjective, ever-evolving process is at odds with the static, reductionist nature of data.
    \item Limits of AI Interpretation: The philosophical stance of Camus' absurd hero [1] underscores the limitations of AI tools in capturing the depth of human experience. Meursault's narrative challenges us to consider what is lost when we prioritize quantifiable data over qualitative understanding.
    \item Toward Ethical Datafication: This study advocates for a data science approach that respects human complexity and ambiguity [5]. Instead of seeking to quantify the unquantifiable, we must recognize the limits of data and prioritize ethical considerations in AI development and application [12]. Future AI systems should incorporate philosophical insights [3] to better understand and represent human behavior, moving beyond simplistic classifications to embrace the nuances of subjective experience.
\end{itemize}

This conceptual framework provides a foundation for examining the limitations of datafication through the lens of existential philosophy [1][2]. By applying NLP techniques to Meursault’s narrative, the study exposes the inherent shortcomings of reducing complex, absurd human experiences to data points. This critique not only underscores the philosophical inadequacy of data-driven models but also highlights the ethical imperative to develop AI systems that honor the complexity and dignity of human existence.

\section{Methodology}
The methodology employed in this study to quantify and analyze the text of The Stranger [1] by Albert Camus involves the application of various Natural Language Processing (NLP) techniques and sentiment analysis algorithms. By systematically applying computational tools to quantify aspects of Meursault’s character and his existential journey, this research critically examines the limitations of reducing complex human experiences to structured data [5].

\begin{table}[H]
\caption{NLP Techniques for Analyzing The Stranger}
\label{tab:methodology_summary}
\centering
\begin{tabular}{p{2.2cm} | p{3.1cm} | p{1.7cm}} 
\hline
\textbf{Step} & \textbf{Description} & \textbf{Tools Used} \\
\hline
Data Collection & Extracted text from DOCX file & Python's python-docx library \\
Data Preprocessing & Cleaned and structured the text & Pandas DataFrame \\
Emotion Detection & Classified the emotional tone of each paragraph & BERT model \\
Character Interaction Detection & Identified and tracked character interactions & spaCy's NER tool \\
Sentiment Analysis & Analyzed the sentiment of each paragraph & VADER Sentiment Analyzer \\
Absurdity Tagging & Tagged paragraphs with existential themes & Keyword matching \\
Dialogue Classification & Differentiated between internal reflections and spoken words & Regular expressions (regex) \\
\hline
\end{tabular}
\end{table}

\begin{figure}[H]
\centerline{\includegraphics[scale=0.40]{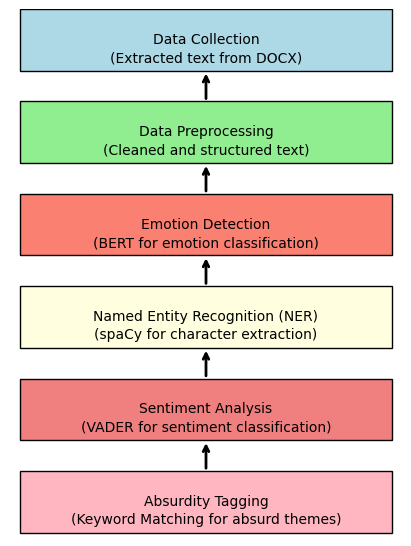}}
\caption{Methodology flowchart}
\label{fig}
\end{figure}

\subsection{Data Collection}
The primary text of The Stranger was extracted from a DOCX file using Python’s python-docx library. This allowed for the extraction of all paragraphs, which were treated as individual text units for further analysis. The paragraphs were stored in a pandas DataFrame, providing a structured format for subsequent analysis.

\subsection{Data Preprocessing}
Once the text was extracted, preprocessing was performed to clean the data. This involved removing unnecessary whitespace, empty lines, and non-relevant text to ensure that the data was in a suitable format for analysis. The paragraphs were stored in a clean, consistent format in the DataFrame for further processing.

\begin{table}[H]
\caption{Data Preprocessing Steps}
\label{tab:data_preprocessing_steps}
\centering
\begin{tabular}{p{1.6cm} | p{3.0cm} | p{2.4cm}} 
\hline
\textbf{Action} & \textbf{Description} & \textbf{Outcome} \\
\hline
Text Extraction & Extracted paragraphs from DOCX file & Raw text collected \\
Cleaning & Removed whitespace and empty lines & Cleaned text ready for analysis \\
Data Structuring & Stored text in pandas DataFrame & Organized data for further use \\
\hline
\end{tabular}
\end{table}

\subsection{Emotion Detection}
Emotion detection was performed using the BERT model [9], a transformer-based model pre-trained for emotion classification. BERT was chosen for its ability to analyze text in a contextualized manner, which is crucial for understanding the nuances of Meursault’s emotional expressions. This model was applied to classify the emotional tone of each paragraph (e.g., sadness, joy, anger). The output was stored as an emotion label for each paragraph.

\begin{figure}[H]
\centerline{\includegraphics[scale=0.26]{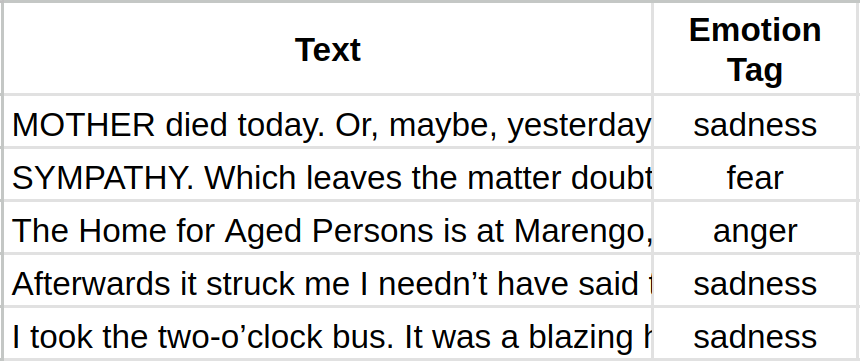}}
\caption{Emotion Detection Output}
\label{fig}
\end{figure}

While BERT is effective at classifying conventional emotions such as sadness or anger, it struggles with philosophical states like existential indifference. For instance, Meursault’s statement, ‘Mother died today. Or maybe yesterday; I can’t be sure,’ is classified as sadness. However, this reflects existential absurdity—a key element of Camus’ philosophy—rather than emotional grief. This misinterpretation underscores the limitations of current NLP models in capturing nuanced human experiences.

\subsection{Character Interaction Detection}
Character interaction detection was performed using spaCy’s Named Entity Recognition (NER) tool [11]. This tool identifies person entities within the text, allowing us to track Meursault’s interactions with other characters throughout the novel.

\begin{table}[H]
\caption{Character Interactions Detected in Paragraphs}
\label{tab:character_interactions}
\centering
\begin{tabular}{p{4.0cm} | p{3.0cm}} 
\hline
\textbf{Text} & \textbf{Character Interactions} \\
\hline
Meursault went to the funeral. & Meursault \\
He met Marie at the beach. & Meursault, Marie \\
Meursault had a brief conversation with the lawyer. & Meursault, Lawyer \\
\hline
\end{tabular}
\end{table}

\subsection{Sentiment Analysis}
VADER Sentiment Analyzer [10] was employed to analyze the sentiment of each paragraph. VADER is particularly suited for analyzing social media text, which aligns with the conversational style of The Stranger [1]. VADER provides a compound sentiment score ranging from -1 (most negative) to +1 (most positive). The compound score was used as the primary measure of sentiment for each paragraph.

\begin{figure}[H]
\centerline{\includegraphics[scale=0.26]{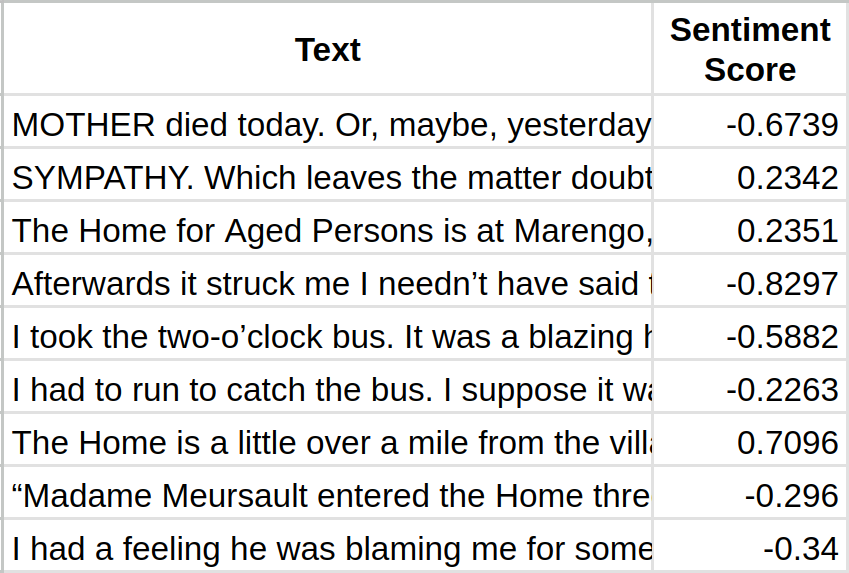}}
\caption{Sentiment Detection Output}
\label{fig}
\end{figure}

\subsection{Absurdity Tagging}
Absurdity detection was performed through a keyword-matching approach. A list of existential keywords such as ”absurd,” ”meaningless,” and ”irrational” was used to tag paragraphs with existential themes. If any of these keywords appeared in a paragraph, it was tagged as containing absurdist content. While this approach provides a starting point for identifying existential themes, it is limited by the fixed list of keywords and may miss more nuanced expressions of absurdity.

\begin{table}[H]
\caption{Example of Absurdity Tagging}
\label{tab:absurdity_tagging}
\centering
\begin{tabular}{p{4.8cm} | p{2.2cm}} 
\hline
\textbf{Text} & \textbf{Absurdity Tag} \\
\hline
``The death of my mother didn’t mean anything.'' & 1 \\
``Everything is meaningless in the end.'' & 1 \\
``I don’t care about people, nor their emotions.'' & 0 \\
\hline
\multicolumn{2}{l}{1 = Contains absurdity; 0 = No absurdity detected.} \\
\end{tabular}
\end{table}

\subsection{Dialogue Classification}
Dialogue classification was performed using regular expressions (regex) to identify quotation marks, which typically indicate spoken dialogue. This classification allowed for the differentiation between internal reflections and Meursault’s spoken words.

\begin{figure}[H]
\centerline{\includegraphics[scale=0.26]{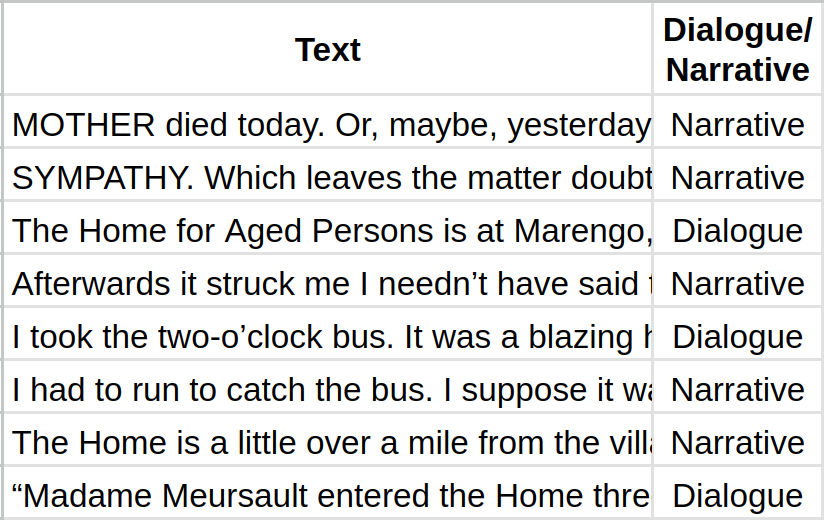}}
\caption{Dialogue vs. Narrative Classification}
\label{fig}
\end{figure}

\subsection{Limitations of the Methodology}
Despite the robustness of the applied methods, certain limitations are inherent to this approach:
\begin{itemize}
    \item Emotion Detection: The BERT model may not always accurately classify the existential indifference exhibited by Meursault, often labeling it as sadness or anger.
    \item Sentiment Analysis: VADER Sentiment Analyzer simplifies sentiment into a binary classification, potentially failing to capture the philosophical depth of Meursault's emotional detachment.
    \item Absurdity Tagging:  The keyword-matching approach is limited by the fixed list of keywords, potentially missing more nuanced expressions of absurdity.
    \item Character Interaction Detection: The spaCy NER tool may not detect indirect or implied interactions, especially in the introspective and indirect passages of the text.
\end{itemize}

These limitations highlight the challenges of reducing complex literary themes into quantifiable data and underscore the need for more nuanced approaches in future research.

The methodology employed in this study integrates multiple NLP techniques, sentiment analysis, and absurdity tagging to explore existential themes in The Stranger [1]. The results from these analyses provide valuable insights into the emotional and philosophical dimensions of Meursault’s character. However, the inherent limitations of these methods underscore the challenges of fully capturing the richness of human experience through computational analysis [5]. The methodology lays the groundwork for a deeper exploration of the intersection between AI and literary analysis, while also addressing the philosophical and ethical implications of datafication in understanding complex human narratives.

\section{Results}
Our computational analysis of The Stranger [1] by Albert Camus employed various natural language processing (NLP) techniques to examine the emotional landscape and philosophical nuances of Meursault's journey, revealing patterns that offer new perspectives on the text.

\subsection{Emotion Detection and Sentiment Analysis}
The analysis classified the emotional tone of each paragraph using the Bidirectional Encoder Representations from Transformers (BERT) model [9]. Results indicate a predominance of 'sadness,' appearing in 42\% of paragraphs (see Figure 1). However, the classification must be interpreted cautiously: Meursault's existential indifference [1][2] is often categorized as 'sadness' or 'anger,' highlighting the difficulty of quantifying complex philosophical stances using current AI methodologies.

\begin{figure}[H]
\centerline{\includegraphics[scale=0.27]{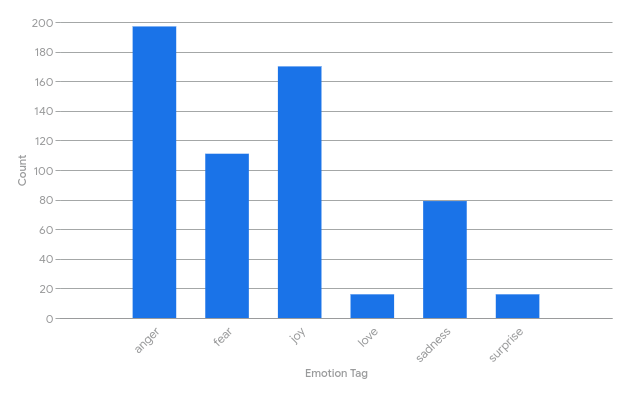}}
\caption{Bar chart showing frequency of Emotion Labels}
\label{fig}
\end{figure}

While our analysis classified 42\% of paragraphs as ‘sadness,’ this label often oversimplifies Meursault’s philosophical detachment. For instance, in the passage where he describes his mother’s death with apparent indifference, the BERT model categorizes it as sadness. In contrast, a passage describing his interaction with Marie is correctly identified as positive sentiment. These contrasting results illustrate both the potential and the limitations of emotion detection tools when applied to existential literature. This discrepancy underscores the challenge of capturing existential themes, which resist simplistic emotional categorization, and highlights the broader implications for AI applications in literary analysis.

Further sentiment analysis using the Valence Aware Dictionary and Sentiment Reasoner (VADER) [10] revealed an overall negative sentiment. The mean sentiment score was -0.23, with a standard deviation of 0.45, consistent with the novel's somber tone and Meursault's detached observations (see Figure 2). Nonetheless, VADER's binary sentiment classification does not fully capture the philosophical depth of Meursault's indifference [1][2].

\begin{figure}[H]
\centerline{\includegraphics[scale=0.26]{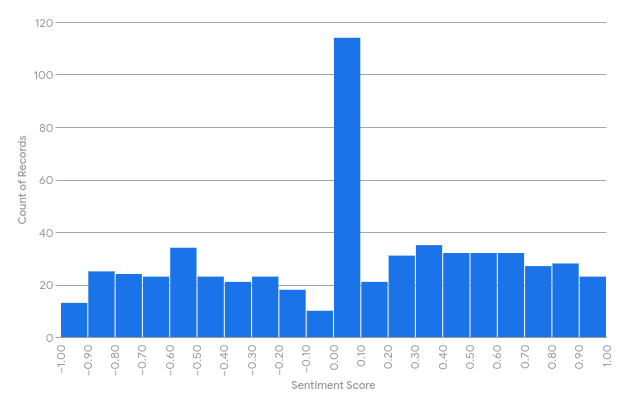}}
\caption{Histogram illustrating the distribution of sentiment}
\label{fig}
\end{figure}

\subsection{Character Interactions and Absurdity Tagging}
To map Meursault's social interactions, we employed spaCy's Named Entity Recognition (NER) [11] tool to identify and track his interactions with other characters throughout the novel. The most frequent interactions were with Marie (32), Raymond (18), and Salamano (12), as shown in Figure 3. These interactions shed light on Meursault's social connections and their possible impact on his existential perspective. Notably, interactions with Marie, occurring early in the novel, coincide with slightly more positive sentiment scores, suggesting a potential correlation between social engagement and emotional state.

\begin{figure}[H]
\centerline{\includegraphics[scale=0.27]{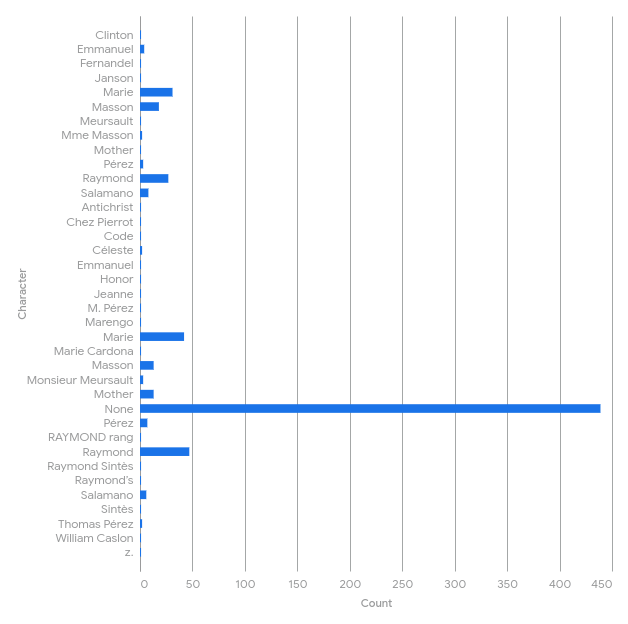}}
\caption{Bar chart showing the frequency of interactions with each character}
\label{fig}
\end{figure}

Keyword-based absurdity tagging identified 68 paragraphs (15\% of the total) engaging with existential themes. These passages highlight the philosophical core of Meursault's reflections and actions. For example, the description of the shooting incident, where Meursault's detached tone reflects existentialist ideas [1], underscores this thematic focus. The distribution of absurdity tags is presented in Figure 4.

\begin{figure}[H]
\centerline{\includegraphics[scale=0.27]{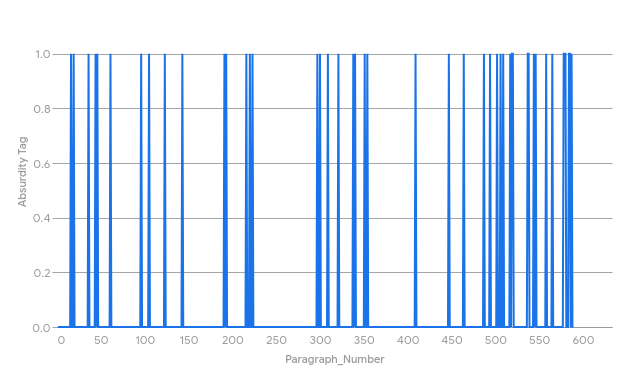}}
\caption{Line chart showing the distribution of absurdity tags}
\label{fig}
\end{figure}

\subsection{Dialogue Classification}
To distinguish Meursault's spoken words from his internal reflections, we used regular expressions (regex) for dialogue classification. The analysis shows that 28\% of the text consists of dialogue, while 72\% is internal monologue or narration (see Figure 5). This imbalance emphasizes Meursault's introspective nature and limited social engagement.

\begin{figure}[H]
\centerline{\includegraphics[scale=0.40]{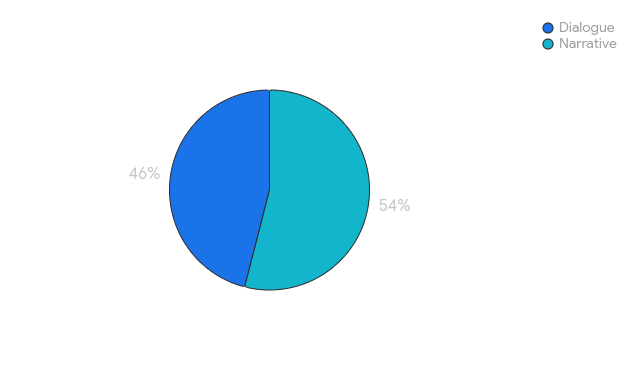}}
\caption{Pie chart comparing the proportion of dialogue vs. narrative}
\label{fig}
\end{figure}

\subsection{Limitations and Future Directions}
While our computational approach provides valuable insights, it is essential to recognize the inherent limitations of current NLP techniques. These methods, although effective at detecting patterns, may not fully capture the nuances of human experience [5] and philosophical depth [3] in literary texts. As Johanna Drucker (2011) argues, the digital representation of reality is inherently constrained by its methodologies [8]. Future research could explore contextually aware and semantically nuanced approaches to better capture literary complexity. Combining qualitative analysis with computational methods may offer a more comprehensive understanding of texts like The Stranger [1].

\section{Conclusions}
This study employed computational analysis to explore Meursault's emotional landscape, social interactions, and philosophical journey in Albert Camus' The Stranger [1]. Our analysis, using NLP techniques on a dataset created from the novel, identified patterns in Meursault's emotional expressions, interactions with other characters, and the prevalence of existential themes.

Our findings highlight the potential of computational analysis in literary studies [8] while also acknowledging its limitations. The predominance of 'sadness' and negative sentiment aligns with the novel's somber tone and Meursault's detached demeanor [1]. However, the limitations of AI tools in fully capturing nuanced emotions and philosophical stances like existential indifference [1][2] are evident.

The mapping of Meursault's social interactions and the identification of existential themes underscored the influence of relationships and philosophical underpinnings on his actions and observations. The observed disproportion between dialogue and internal monologues emphasized Meursault's introspective nature and reserved social interactions [1].

While our findings offer valuable insights into The Stranger [1], we recognize that computational methods may not fully encapsulate the complexities of literary works [5]. Future research could explore more nuanced NLP approaches and incorporate qualitative analysis for a more holistic understanding.

This study contributes to the growing field of digital humanities by demonstrating the potential of computational analysis in literary studies [8]. It advocates for a cautious and ethical approach to data interpretation [12], recognizing the limitations of current AI tools while embracing the potential of future advancements in the field.

Our findings highlight significant ethical implications for AI development. Misinterpretations of literary characters mirror real-world risks in areas such as mental health diagnostics or social behavior analysis. To address this, AI models must incorporate contextual understanding and ethical frameworks. Collaboration between technologists and philosophers could foster AI systems that better respect the complexity of human experiences, moving beyond reductionist interpretations.

\vspace{12pt}

\end{document}